 \documentclass[final,3p,times,numbers]{elsarticle}

\usepackage{amssymb}
\usepackage{lipsum}

\biboptions{sort&compress}

\usepackage{tikz}
\usetikzlibrary{patterns}
\usetikzlibrary{positioning,arrows}
\usetikzlibrary{decorations.pathmorphing}
\usetikzlibrary{decorations.markings}
\usetikzlibrary{snakes}
\usetikzlibrary{matrix}

\usepackage{amsmath}

\usepackage{amsfonts}
\usepackage{enumerate}
\usepackage{mathrsfs}
\usepackage{tensor}

\def\be{\begin{equation}}
	\def\ee{\end{equation}}
\def\bea{\begin{eqnarray}}
	\def\eea{\end{eqnarray}}
\newcommand{\nn}{\nonumber}

\def\apjl{\ref@jnl{ApJ}}

\journal{Nuclear Physics B}

\begin{document}

\begin{frontmatter}



\title{Exact expressions for the 5-Point Liouville conformal block with a level-two degenerate field insertion}


\author[first]{Hasmik Poghosyan}
\author[first]{Rubik Poghossian}
\affiliation[first]{organization={Yerevan Physics Institute},
            addressline={Alikhanian Br. 2}, 
            city={ Yerevan},
            postcode={0036}, 
            country={Armenia}}

\begin{abstract}
In this paper we investigate the  5-point Liouville conformal block 
with a level two degenerate field insertion. Our main tool is the BPZ differential equation, which, 
upon placing three of the insertions at the standard positions $\infty$, $1$, and $0$, 
reduces to a linear differential equation which is of order two in the degenerate 
insertion point $z$, and order one in the remaining point $x$. In a previous paper, it was conjectured that the solution could be expressed in terms of a single hypergeometric function and its derivative, with coefficients computable via recursive relations up to the desired order $x^k$. In this paper, we simplify these recursion relations and provide a rigorous inductive proof of the conjecture. 

A key advantage of our representation of the 5-point conformal block is that the hypergeometric connection formulae readily relate its different regions of analyticity.

In the quasi-classical limit, the 5-point BPZ equation reduces to the Heun equation. Consequently, we recover a recently proposed representation of the Heun equation in terms of a single hypergeometric function, which has proven to be highly effective in the analysis of gravitational perturbations of black holes.	
\end{abstract}




\end{frontmatter}




\section{Introduction}
\label{introduction}

	In two-dimensional conformal field theories (CFTs), multipoint correlation functions with at least one degenerate field insertion satisfy the Belavin–Polyakov–Zamolodchikov (BPZ) differential equation \cite{Belavin:1984vu}. For four-point functions, the BPZ equation reduces to an ordinary linear differential equation in a single variable: the cross ratio of the insertion points. The study of this equation, along with its analogues in theories with extended conformal symmetries, has proven to be highly effective for the explicit computation of OPE structure constants. Notable examples include the $\mathcal{N}=1$ super-minimal models \cite{Zamolodchikov:1988nm}, the parafermionic series with spin-$4/3$ currents \cite{Poghossian:1988yao}, Liouville theory \cite{Teschner:1995yf}, and $\mathcal{N}=1$ super-Liouville theory \cite{Poghossian:1996agj}.
For 5-point functions, the BPZ equations become linear partial differential equations (PDEs) in two independent variables, corresponding to the two cross ratios of the insertion points. 

In principle, 5-point conformal blocks can be computed order by order as a power series in the ratios of insertion points using the conformal Ward identities \cite{Belavin:1984vu}. However, this method becomes impractical for solving the differential equations of our interest, particularly when higher-order terms in the expansion are required.

The discovery of the AGT (Alday–Gaiotto–Tachikawa) correspondence \cite{Gaiotto:2009we,Alday:2009aq} has opened new avenues for addressing the BPZ differential equation. According to AGT, two-dimensional CFT correlation functions are intimately related to the partition functions of four-dimensional $\mathcal{N}=2$ supersymmetric gauge theories in the $\Omega$-background. This correspondence enables the use of powerful localization techniques \cite{Nekrasov:2002qd,Flume:2002az,Bruzzo:2002xf,Nekrasov:2003rj,Flume:2004rp} to compute conformal blocks efficiently.

In the later work \cite{Gaiotto:2012sf}, a new method was proposed for computing the 5-point conformal block with a level-two degenerate field insertion. It was observed that the solution to the BPZ differential equation can be expressed as a double series whose coefficients satisfy a recursive relation. Furthermore, it was argued that this double sum could be recast in terms of differential operators acting on a hypergeometric function, thus making the analytic structure of the series more accessible.

This idea was further developed in \cite{Poghosyan:2025rhj}, where it was conjectured that the differential operator in question can be chosen to be of first order. An efficient recursion relation for the coefficients of this operator was formulated and verified to hold up to fairly high orders in the series expansion.

In the present paper, we revisit the recursion relation introduced in \cite{Poghosyan:2025rhj} and uncover several underlying symmetries. These symmetries allow us to reformulate the recursion relation in a significantly simplified form. This new representation enables us to provide a rigorous proof of the conjecture originally proposed in \cite{Poghosyan:2025rhj}.

We anticipate that our results can be extended to the class of  irregular blocks, which attract  attention 
due to their relevance in Argyres Douglas theories \cite{Argyres:1995jj,Argyres:1995xn} in $\Omega$-background 
\cite{Gaiotto:2009ma,Gaiotto:2012sf,Nishinaka:2012kn,
	Bonelli:2016qwg,Nishinaka:2019nuy,
	Poghosyan:2023zvy,Poghossian:2025nef,Iorgov:2025hxt}. 

Let us also highlight several directions where our results may prove to be relevant.

One such direction emerges when the $\Omega$-background parameters are constrained to the self dual case $\epsilon_1 = -\epsilon_2$. In this regime, the Nekrasov partition functions are closely related to the $\tau$-functions of Painlev\'{e} transcendents \cite{Gamayun:2013auu,Bonelli:2016qwg,Poghosyan:2023zvy,Bonelli:2025owb}.

Another direction involves  the so-called Nekrasov–Shatashvili (NS)  limit, defined by setting $\epsilon_2=0$    \cite{Nekrasov:2009rc} (see also \cite{Poghossian:2010pn,Fucito:2011pn}).
This limit has found various applications sometimes in subjects far away from each other, e.g. in 
relation to Painlev\'{e} $VI$ \cite{Litvinov:2013sxa} and the analysis of gravitational perturbations
\cite{Aminov:2020yma,Bianchi:2021xpr,Bianchi:2021mft,Fioravanti:2021dce,
	Consoli:2022eey,Fucito:2023afe,Aminov:2023jve,Fucito:2024wlg,DiRusso:2024hmd,Bautista:2023sdf,Liu:2024eut,Cipriani:2025ikx}.

The paper is organized as follows:

In Section~\ref{5pntBlock}, we provide a brief review of two-dimensional Liouville field theory. In particular, we present the BPZ differential equation satisfied by the 5-point conformal block with a level-two degenerate field insertion.

Section~\ref{RevRec} is a discussion on the recursive approach for computing the 5-point conformal block.

In Section~\ref{New_rec}, we introduce a new, compact formulation of this recursion relation. 
This alternative representation was essential for the proof of the conjecture.

Finally, in Section~\ref{theProof}, we present the detailed proof of the conjecture.

\section{5-point Liouville conformal block with a degenerate field insertion}
\label{5pntBlock}
Our conventions in this paper are the same as in \cite{Poghosyan:2025rhj}. 
Here, we provide a short introduction to them.
In two-dimensional Liouville field theory, the Virasoro central charge is parametrized as
\bea
c=1+6Q^2
\\
Q=b+b^{-1}
\eea
where $b$ is the dimensionless coupling constant of the theory. The conformal dimensions
$h_p$ of primary fields or states are related to their momentum parameters via
\bea
h_p=\frac{Q^2}{4}-p^2
\eea
\begin{figure}[t]
	\begin{center}
		\begin{tikzpicture}[scale=0.50]
			\draw (1,1)	-- (3,1);
			\draw (3,1)	-- (5,1);
			\draw (5,1)	-- (7,1);
			\draw (7,1)	-- (9,1);
			\draw (3,1) -- (3,3);
			\draw[dashed] (5,1) -- (5,3);
			\draw (7,1) -- (7,3);
			\node[scale=0.8] at (0.5,1){$\infty$};
			\node[scale=0.8] at (9.4,1){$0$};
			\node[scale=0.8] at (3,3.4){$k_0$};
			\node[scale=0.8] at (5,3.4){$k_{deg}$};
			\node[scale=0.8] at (7,3.4){$k_2$};
			\node[scale=0.8] at (2,0.45){$p_0$};
			\node[scale=0.79] at (4,0.55){$p^{-\sigma }$};
			\node[scale=0.8] at (6,0.45){$p$};
			\node[scale=0.8] at (8,0.45){$p_3$};
			\node[scale=0.8] at (2.6,2){$1$};
			\node[scale=0.8] at (4.6,2){$z$};
			\node[scale=0.8] at (6.6,2){$x$};
		\end{tikzpicture}
	\end{center}

	\caption{The conformal block ${\cal F}(z,x)$ given in (\ref{5pointCB}).}
	\label{fig:Fusion_Braiding}
\end{figure}
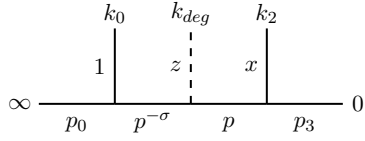
The central object of our interest in this work is the 5-point conformal block (see Fig.~\ref{fig:Fusion_Braiding}) with a level two degenerate field insertion, for which we adopt the following notation:
\bea
\label{5pointCB}
{\cal F}(z,x)=\langle p_0| V_{k_0} (1) V_{k_{deg}}(z) V_{k_2} (x)  |p_3\rangle
\eea
It satisfies the BPZ  differential equation \cite{Belavin:1984vu}
	\bea
	\label{BPZ}
	& \left(
	\frac{1}{b^2}\frac{\partial ^2}{\partial z^2}-\frac{2 z-1}{z (z-1)} \frac{\partial }{\partial z}+
	\frac{x (x-1) }{z (z-1) (z-x)}\frac{\partial }{\partial x}+
	\left(\frac{h_{k_2}}{(z-x)^2}+\frac{h_{p_3}}{z^2}+\frac{h_{k_0}}{(z-1)^2}-\frac{\delta }{z (z-1)}\right)
	\right){\cal F}(z,x)=0
	\eea
where 
\bea
\delta =-h_{p_0}+h_{k_0}+h_{k_{deg}}+h_{k_2}+h_{p_3}
\eea
and 
$V_{k_{deg}}(z)$   denotes the level two degenerate field with momentum parameter
\bea
\label{deg_field}
k_{deg}= b+\frac{1}{2 b}
\eea
OPE fusion rules impose restrictions on the intermediate field dimensions:
\bea
\label{OPEint}
p_1= p-\frac{ \sigma b}{2}\,,
\quad
p_2= p
\quad
{\rm where }
\quad
\sigma=\pm 1
\eea
We will also use  the notation
\bea
p^\sigma= p+ \frac{\sigma b}{2}
\eea	
To simplify the BPZ equation, we factor out the free-field contribution by defining a new function 
\bea
\label{GvsF}
G(z,x)=A^{-1}(z,x)	{\cal F}(z,x) 
\eea
with the prefactor
	\bea
	\label{Azx}
	A(z,x)=
	(1-x)^{-\frac{1}{2} \left(Q-2 k_0\right) \left(Q-2 k_2\right)}x^{-\frac{1}{2} \left(Q-2 k_2\right) \left(Q-2 p_3\right)} (1-z)^{\frac{b}{2}  \left(Q-2 k_0\right)} z^{\frac{b}{2}  \left(Q-2 p_3\right)}  (z-x)^{\frac{b}{2}  \left(Q-2 k_2\right)}
	\eea
The function $G(z,x)$  then satisfies the transformed differential equation:
	\bea
	\label{BPZren}
	& \left(
	\frac{\partial ^2}{\partial z^2}
	+\left(\frac{\kappa _2+\kappa _3}{z-1}+\frac{1-\gamma _0+2 \kappa _1}{z-x}+\frac{\gamma _0}{z}\right)\frac{\partial }{\partial z}
	+\frac{b^2 x (x-1)}{z (z-1) (z-x)}\frac{\partial }{\partial x}
	+ \frac{\left(\kappa _1+\kappa _2\right) \left(\kappa _1+\kappa _3\right)}{z (z-1) }
	\right)G(z,x)=0\qquad
	\eea
with
\bea
\label{kappaToP}
\gamma _0 &=& 1-2 b p_3 \,,
\qquad
\kappa _1 = \frac{b}{2} \left(Q-2  k_2-2  p_3\right)
\\
\kappa _2 &=&  \frac{1}{2} \left(1-2 b k_0-2 b p_0\right)\,,
\qquad
\kappa _3 = \frac{1}{2} \left(1-2 b k_0+2 b p_0\right)
\eea
\section{Solution in terms of a hypergeometric function}
\label{RevRec}
In \cite{Poghosyan:2025rhj}, a recursion relation was derived that enables the conformal block \( G(z,x) \) to be constructed order by order as a power series in $ x $. 
However, this method relies on a conjectural assumption, which we aim to establish rigorously in the present paper. 
To formulate the conjecture precisely, let us briefly review the recursive construction of the conformal block. 
The solution of the differential equation~\eqref{BPZren} can consistently be represented in the following form:
\bea
\label{Ganz1}
&&G(z,x|p^{-\sigma},p)=x^{\frac{\kappa _1^2}{b^2}-p^2}(P(x,z)H_\sigma(z)
+\hat{P}(x,z)z H^{'}  _{\sigma}   (z))
\eea
where
\bea
\label{H1}
&&H_\sigma(z)
=
z^{b p \sigma -\kappa _1} \, _2F_1\left(b p \sigma +\kappa _2,b p \sigma +\kappa _3;2 b p \sigma +1;z\right)
\eea
and 
\bea
\label{cij}
P(x,z)=\sum_{i=0}^\infty \sum_{j=0}^{\infty}h_{i,j}x^i z^{j-i}\,,
\quad
\hat{P}(x,z)=\sum_{i=0}^\infty \sum_{j=0}^{\infty}\hat{h}_{i,j}x^i z^{j-i}
\eea 
Actually as we will prove in  Section~\ref{theProof}, \( h_{i,j} = \hat{h}_{i,j} = 0 \) for \( j > i \), so that
(\ref{cij}) can be rewritten as 
\bea
\label{Pfin}
P(x,z) = \sum_{i=0}^{\infty} \sum_{j=0}^{i} h_{i,j}\, x^i z^{j-i},
\qquad
\hat{P}(x,z) = \sum_{i=0}^{\infty} \sum_{j=0}^{i} \hat{h}_{i,j}\, x^i z^{j-i}.
\eea
Consequently, at any fixed order in \( x \), say \( x^k \), both \( P \) and \( \hat{P} \) reduce to finite polynomials in \( z^{-1} \).
Notice that we have adopted a more detailed notation $G(z,x|p_1,p_2)$ for the function $G(z,x)$, explicitly 
indicating intermediate state momenta $p_1$ and $p_2$.
By imposing differential equation (\ref{BPZren}) on (\ref{Ganz1}) and 
using the identity
\bea 
&H_{\sigma }^{''}(z)=
\frac{1+2 \kappa _1-\left(2 \kappa _1+\kappa _2+\kappa _3+1\right) z}{(z-1) z} H_{\sigma }^{'}(z)
-\frac{b^2 p^2-\kappa _1^2+\left(\kappa _1+\kappa _2\right) \left(\kappa _1+\kappa _3\right) z}{(z-1) z^2}H_{\sigma }(z)
\eea 
to reduce higher order derivatives to $H_\sigma (z)$ and $H_{\sigma }^{'}(z)$, 
recursion relations were found. These take a much simpler form when one introduces
\bea
h_{i,j}=\sum_{\sigma=\pm 1}\left(\kappa _1+ \sigma b  p\right) t_{i,j}(\sigma p)\,,
\quad
\hat{h}_{i,j} = \sum_{\sigma=\pm 1} t_{i,j}(\sigma p)
\eea
Then the recursion relations for coefficients $t_{i,j}(\sigma p)$ have the form
\begin{figure}[t!]
	\centering
	\begin{tikzpicture}[scale=0.50]
		\draw[step=1cm,gray,dotted,very thin] (-0.5,-0.8) grid (10,6.6);
		\path[pattern={north east lines},pattern color= lightgray]  (3,4) --  (3,0.9)-- (6,0.9)-- (6,1);
		\draw (6.2,0.8)--(3,4);
		\draw[->] (0.1,4)--(8,4) node[right]{$j$};
		\draw[<-] (3,0.3)--(3,5.7) ;
		\node at (3,-0.2){i};
		\draw[thick, color=gray]  (1,4)--(3,4)--(3,5)--(1,5)--(1,4);
		\foreach \p in {(1,4),(2,4),(1,5),(2,5), (3,5) }
		\fill[orange] \p circle(.1);
		\fill[blue]  (3,4)  circle(.16);
		\fill[orange]  (3,4)  circle(.12);
		\foreach \p in {(3,1),(3,2),(3,3),(4,1),(4,2),(4,3),(5,1),(5,2),(6,1)}
		\fill[blue] \p circle(.12);
		\foreach \p in {(6,2),(5,3),(6,3),(4,4),(5,4),(6,4)}
		\fill[gray] \p circle(.12);
	\end{tikzpicture}
	\caption{It is proven in Section~\ref{theProof} that at the dots outside the shaded region $t_{i,j}=0$. 
		The values $t_{i,0}$ on the  vertical  boundary are given in (\ref{ti0}).}
	\label{fig:tRec}
\end{figure}
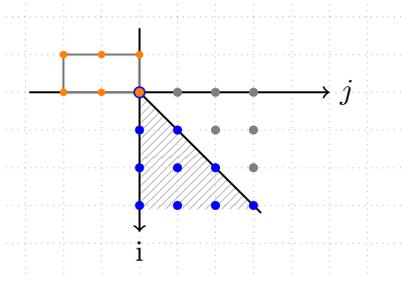
\bea
\label{RecrelTr0}
B_{(0,0)(i,j)}(p)t_{i,j}(p)=
\sum_{\substack{s \in \{0,1\},l\in{\{0,1,2\}} \\ (s,l)\neq (0,0)}} 
B_{(s,l)(i,j)}(p)t_{i-s,j-l}(p)
+\hat{B}_{(s,l)(i,j)}(p)t_{i-s,j-l}(-p)
\eea
where
\bea
\label{B_Bhat}
B_{(0,0)(i,j)}(p)&=&4 b^2 p^2 (j-i)+2 b p \left(b^2 i+(i-j)^2\right)
\\
B_{(1,0)(i,j)}(p)&=&2 b p \left(b p-i+j-\kappa _1+1\right) \left(b p+\gamma _0-i+j-\kappa _1\right)
\eea
\bea
B_{(0,1)(i,j)}(p)=b p \left(2 i \left(b^2-4 j+4\right)+4 i^2+4 (j-1)^2-\kappa _2-\kappa _3\right)
-b^2 p^2 (6 (i-j)+7)+\kappa _2 \kappa _3 (2 (i-j)+1)
\eea
\bea
\label{B02}
B_{(0,2)(i,j)}(p)=2 (i-j+2) \left(b p (b p-i+j-2)-\kappa _2 \kappa _3\right)
\eea
	\bea
	B_{(1,1)(i,j)}(p)=-4 b^3 p^3	+b^2 p^2 \left(-3 \gamma _0+6 (i-j)+6 \kappa _1+4\right)+ b \gamma _0 p \left(4 (i-j)+4 \kappa _1+\kappa _2+\kappa _3\right)
	\nn\\
	+b p \left(2 b^2 (i-1)-2 \kappa _1 \left(4 (i-j)+2 \kappa _1+\kappa _2+\kappa _3+2\right)-4 (i-j) (i-j+1)\right)
	+\kappa _2 \kappa _3 \left(\gamma _0-2 \left(i-j+\kappa _1\right)\right)
	\quad
	\eea
	\bea
	B_{(1,2)(i,j)}(p)=	2 b^3 p^3+b^2 p^2 \left(\gamma _0-2 i+2 j-2 \kappa _1-3\right)-b \gamma _0 p \left(2 i-2 j+2 \kappa _1+\kappa _2+\kappa _3+2\right)
	\nn\\
	+b p \left(2 \kappa _1 \left(2( i- j)+\kappa _1+\kappa _2+\kappa _3+3\right)-2 b^2 (i-1)+2 (i-j+1) (i-j+2)+\kappa _2+\kappa _3\right)
	\nn\\
	+\kappa _2 \kappa _3 \left(-\gamma _0+2 i-2 j+2 \kappa _1+3\right)\quad
	\eea
\bea
\hat{B}_{(1,0)(i,j)}(p)&=&0
\\
\hat{B}_{(0,1)(i,j)}(p)&=&(2 (i-j)+1) \left(b p+\kappa _2\right) \left(b p+\kappa _3\right)
\\
\label{Bh02}
\hat{B}_{(0,2)(i,j)}(p)&=&-2 (i-j+2) \left(b p+\kappa _2\right) \left(b p+\kappa _3\right)
\\
\hat{B}_{(1,1)(i,j)}(p)&=&\left(b p+\kappa _2\right) \left(b p+\kappa _3\right) \left(\gamma _0-2 \left(i-j+\kappa _1\right)\right)
\\
\hat{B}_{(1,2)(i,j)}(p)&=&\left(b p+\kappa _2\right) \left(b p+\kappa _3\right) \left(3-\gamma _0+2( i- j+ \kappa _1)\right)
\eea
These relations together with initial conditions
\bea
\label{ini_cond_00}
&&t_{0,0}( p)=\frac{1}{2 b  p} \\
\label{ini_cond_neg}
&&t_{i,j}=0\,,\quad {\rm if} \quad i<0 \quad {\rm  or}\quad
j<0 
\eea
uniquely determine the coefficients $t_{i,j}$ for all $i\ge 0,\,j\ge 0$. 
In particular through several iterations we get
\bea
\label{ti0}
t_{i,0}(p)=\frac{\left(\kappa _1-b p\right)_i \left(1-b p-\gamma _0+\kappa _1\right)_i}
{2 b p\left(b Q-2 p b\right)_i i! }
\eea
Let us emphasize that, in deriving the recursion relation~\eqref{RecrelTr0}, 
we performed shifts in the indices \( i \) and \( j \). 
Such manipulations are justified provided  the double summation in~\eqref{cij} 
runs over non-negative integers \( i,j \ge 0 \). 

By iterating the recursion relation together with the initial condition~\eqref{ini_cond_00}, 
one can compute the coefficients \( h_{i,j} \) explicitly for various values of \( i \) and \( j \). 
From these computations, one observes that whenever \( j > i \), the coefficients vanish,
\[
h_{i,j} = \hat{h}_{i,j} = 0,
\]
which is equivalent to \( t_{i,j} = 0 \). 

In our previous paper~\cite{Poghosyan:2025rhj}, we verified this property for sufficiently large values of the indices \( i \) and \( j \). 
However, a rigorous proof was not provided. 
The main purpose of the present paper is to fill this gap. 
In the next section, we derive several consequences of the recursion relation~\eqref{RecrelTr0} that will ultimately lead to the desired result.
\section{Equivalent expressions for the recursion relation}
\label{New_rec}
Before proving the conjecture, let us first rewrite the recursion relation in a more elegant form.
We will then derive an equivalent recursion relation, which will serve as the foundation for our proof.

By analyzing the structure of equation (\ref{RecrelTr0}), we observe that it can be represented as
\bea
\label{recRelMidle}
f_{i,j}(p)+r_{i,j}( p)-(f_{i,j-1}( p)+r_{i-1,j}( p))=0
\eea
where
\bea
r_{i,j}(p)=-\left(b p+\kappa _2\right) \left(b p+\kappa _3\right) \left(t_{i,j-1}(p)+t_{i,j-1}(-p)\right)
\eea
and
\bea
\nn
f_{i,j}(p)&=&C_{0,0}(p)  t_{i,j}(p)+C_{0,1}(p)t_{i,j-1}(p)+ C_{1,0}(p) t_{i-1,j}(p)
\\
\label{fij}
&+&C_{1,1}(p)t_{i-1,j-1}(p)
+\hat{C}_{1,1}(p)t_{i-1,j-1}(-p)+
\hat{C}_{0,1}(p) t_{i,j-1}(-p)
\eea
with
\bea
C_{0,0}(p)&=&-2 b p \left(b^2 i+2 b p (j-i)+(i-j)^2\right)
\\
C_{0,1}(p)&=&2 (i-j+1) \left(b p (-b p+i-j+1)+\kappa _2 \kappa _3\right)
\\
C_{1,0}(p)&=&2 b p \left(b p-i+j-\kappa _1+1\right) \left(b p+\gamma _0-i+j-\kappa _1\right)
\eea
\bea
C_{1,1}(p)&=&
\kappa _3 \left(\left(\gamma _0-2 \kappa _1\right) \left(b p+\kappa _2\right)-b p+\kappa _2 (-2 i+2 j-1)\right)
\\ \nn
&+& b p \left(2 b^2 \left(i-p^2-1\right)+b p (2 i-2 j+1)-2 (i-j) (i-j+1)\right)
\\ \nn
&+& b p \left(\gamma _0 \left(2 i-2 j+2 \kappa _1+\kappa _2-b p\right)-2 \kappa _1 \left(2 i-2 j+\kappa _1+\kappa _2+1-b p\right)-\kappa _2\right)
\eea
\bea
\hat{C}_{0,1}(p)&=&2 (i-j+1) \left(b p+\kappa _2\right) \left(b p+\kappa _3\right)
\\
\hat{C}_{1,1}(p)&=&-\left(b p+\kappa _2\right) \left(b p+\kappa _3\right) \left(1-\gamma _0+2 i-2 j+2 \kappa _1\right)
\eea
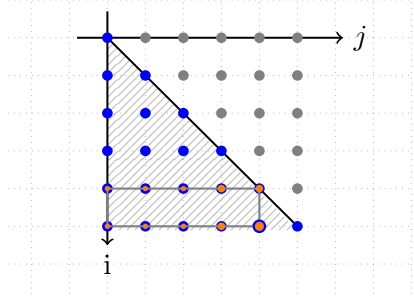
\begin{figure}[t]
	\centering
	\begin{tikzpicture}[scale=0.50]
		\draw[step=1cm,gray,dotted,very thin] (0.4,-2.8) grid (11,5.1);
		\path[pattern={north east lines},pattern color= lightgray]  (3,4) --  (3,-1.1)-- (8,-1.1);
		\draw (8,-1)--(3,4);
		\draw[->] (2.2,4)--(9.2,4) node[right]{$j$};
		\draw[<-] (3,-1.5)--(3,4.7);
		\node at (3,-2){i};
		\foreach \p in {(3,-1),(3,0),(3,1),(3,2),(3,3), (3,4),(4,-1),(4,0),(4,1),(4,2),(4,3),(5,-1),(5,0),(5,1),(5,2),(6,-1),(6,0),(6,1),(7,0),(7,-1),(8,-1)}
		\fill[blue] \p circle(.14);
		\foreach \p in {(6,2),(5,3),(6,3),(4,4),(5,4),(6,4),(7,4),(8,4),(7,3),(8,3),(7,2),(8,2),(7,1),(8,1),(8,0)}
		\fill[gray] \p circle(.14);
		\draw[thick, color=gray]  (3,-1)--(7,-1)--(7,0)--(3,0)--(3,-1);
		\foreach \p in {(6,-1),(7,-1),(6,0),(7,0) }
		\fill[orange] \p circle(.1);
		
		\foreach \p in {(3,-1),(4,-1),(5,-1),(6,-1),(3,0),(4,0) ,(5,0)}
		\fill[orange] \p circle(.07);

		\fill[blue]  (7,-1)  circle(.18);
		\fill[orange]  (7,-1)  circle(.12);
	\end{tikzpicture}
	\caption{The special case $(i,j)=(5,4)$ of (\ref{recRelMain}). The related nodes are indicated in orange.}
	\label{fig:tRecNew}
\end{figure}
From (\ref{recRelMidle})  we derive \footnote{We have used the trivial identity $\sum _{l=1}^j \left(f_{i,l}-f_{i,l-1}\right)=f_{i,j}-f_{i,0}$ and noticed that $f_{i,0}=0$, as follows from (\ref{fij}) and (\ref{ti0}).}
\bea
\label{recRelMain}
f_{i,j}(p)+\sum_{l=1}^j (r_{i,l}(p)-r_{i-1,l}(p)) =0
\eea
This provides another recursion relation through which the functions $t_{i,j}$
can be computed.
A pictorial representation of this recursion is shown in Fig. \ref{fig:tRecNew}.
Summing this relation over the second index gives
\bea
\label{SumRecRelMain}
\sum_{m=0}^j\left( f_{i,m}(p)+\sum_{l=1}^m (r_{i,l}(p)-r_{i-1,l}(p))\right) =0
\eea
Inserting the expression for $f_{i,j}$ from equation (\ref{fij}) and using the boundary expression (\ref{ti0}), we obtain:
	\bea
	\nn
	2 b p  \left(b p-i+j-\kappa _1+1\right) \left(b p+\gamma _0-i+j-\kappa _1\right)t_{i-1,j}(p)-2 b p  \left(b^2 i+2 b p (j-i)+(i-j)^2\right)t_{i,j}(p)
	\\ \nn
	+\sum_{l=0}^{j-1} \bigg( 2 \left((i-l) \left(b^2 p^2+\kappa _2 \kappa _3\right)-b^3 i p\right)\left(t_{i,l}(p)-t_{i-1,l}(p)\right) -2 b p \left(b^2+\kappa _2+\kappa _3\right) t_{i-1,l}(p)
	\\ \nn
	+\left(b p+\kappa _2\right) \left(b p+\kappa _3\right) \left((j-l) \left(t_{i-1,l}(p)-t_{i,l}(p)\right)+(2 i-j-l) \left(t_{i,l}(-p)-t_{i-1,l}(-p)\right)\right)
	\\
	\label{SumEq}
	+\left(\gamma _0-2 \kappa _1+1\right) \left(b p+\kappa _2\right) \left(b p+\kappa _3\right) \left(t_{i-1,l}(p)+t_{i-1,l}(-p)\right)\bigg)=0
	\eea
\section{Proof of the conjecture}
\label{theProof}
Now we are ready to prove the {\bf statement}: $t_{i,j}=0$, provided $j>i$.\\
We will proceed by induction on the index $i$. If $i=0$, (\ref{RecrelTr0}) takes the form
\bea
\label{req_i_0} 
&&2 b j p (2 b p+j) t_{0,j}(p)=\nn\\
&&\qquad 2 (j-2) \left[\left(\kappa _2 \kappa _3-b p (b p+j-2)\right)t_{0,j-2}(p) +\left(b p+\kappa _2\right) \left(b p+\kappa _3\right)
t_{0,j-2}(-p)\right]\nn\\
&&\qquad +\left(b p \left(b p (6 j-7) +4 (j-1)^2-\kappa _3\right)+\kappa _2 \left((1-2 j) \kappa _3-b p\right)\right)t_{0,j-1}(p) \nn\\
&&\qquad +(1-2 j) \left(b p+\kappa _2\right) \left(b p+\kappa _3\right) t_{0,j-1}(-p)\qquad\qquad
\eea 
If $j=0$, the relation (\ref{req_i_0}) is trivially satisfied due to the initial 
condition (\ref{ini_cond_neg}). Considering  the cases $j=1$ and $j=2$ one easily finds 
that $t_{0,1}(p)=t_{0,2}(p)=0$, hence, due to (\ref{req_i_0}), $t_{0,j}(p)=0$ for any $j>2$ 
as well. 

Assume $t_{i',j}(p)=0$ for $i'\in \{0,1,2,\cdots ,i-1\}$ and $j>i'$. Specifying 
$j=i+1$ in (\ref{recRelMain}) and omitting terms which vanish due to our assumption, 
we obtain
\bea
\label{recRelMain_special}
2 b p  \left(b^2 i+2 b p+1\right)t_{i,i+1}(p)=
 \left(b p+\kappa _2\right) \left(b p+\kappa _3\right) \left(\sum _{l=0}^{i-1} \left(t_{i-1,l}(p)+t_{i-1,l}(-p)\right)-\sum _{l=0}^{i}\left(t_{i,l}(p)+t_{i,l}(-p)\right)\right)
\eea 
Similarly, substituting $i\to i'$, $j\to i'+1$ in (\ref{recRelMain}) we get
\bea
\sum _{l=0}^{i'-1} \left(t_{i'-1,l}(p)+t_{i'-1,l}(-p)\right)=\sum _{l=0}^{i'}\left(t_{i',l}(p)+t_{i',l}(-p)\right)
\eea 
Due to  (\ref{ini_cond_00}) $t_{0,0}=\frac{1}{2bp}$, hence, starting from $i'=1$,
and repeatedly increasing $i'$ until reaching $i'=i-1$ we readily establish that  
\bea
\label{cond-sum_t}
\sum _{l=0}^{i'}\left(t_{i',l}(p)+t_{i',l}(-p)\right)=0
\eea 
for $i'\in \{0,1,2, \cdots , i-1\}$.

On the other hand, specializing the l.h.s. of (\ref{SumEq}) to $j=i+1$ and subtracting from 
the resulting expression the same expression with $p$ substituted by  $-p $, we deduce
\bea 
-2bp(1+2bp+b^2i)t_{i,i+1}(p)-2bp(1-2bp+b^2i)t_{i,i+1}(-p)\nn\\+
2 b p \left(b^2 (i-1)+\left(\kappa _2+\kappa _3\right) \left(1+\gamma _0-2 \kappa _1\right)\right)\sum_{l=0}^{i-1}
\left(t_{i-1,l}(p)+t_{i-1,l}(-p)\right)\nn\\
-2 b p \left(b^2 i+\kappa _2+\kappa _3\right)\sum_{l=0}^{i}
\left(t_{i,l}(p)+t_{i,l}(-p)\right)=0
\eea
With the help of  (\ref{recRelMain_special}), the two terms on the first row of the above expression 
can be replaced by
\bea 
-\left(b p+\kappa _2\right) \left(b p+\kappa _3\right) \left(\sum _{l=0}^{i-1} 
\left(t_{i-1,l}(p)+t_{i-1,l}(-p)\right)-\sum _{l=0}^{i}\left(t_{i,l}(p)+t_{i,l}(-p)\right)\right) \nn\\
+\left(-b p+\kappa _2\right) \left(-b p+\kappa _3\right) \left(\sum _{l=0}^{i-1} 
\left(t_{i-1,l}(p)+t_{i-1,l}(-p)\right)-\sum _{l=0}^{i}\left(t_{i,l}(p)+t_{i,l}(-p)\right)\right) 
\eea 
Taking then  into account (\ref{cond-sum_t}), 
we get
\bea 
-2 b^3 i p \sum _{l=0}^{i} 
\left(t_{i,l}(p)+t_{i,l}(-p)\right)=0
\eea  
Thus the equation (\ref{cond-sum_t}) holds for $i'=i$ as well. 

Turning back to (\ref{recRelMain_special}), we see that
\bea 
t_{i,i+1}(p)=0
\eea 
Examining the recursion relation (\ref{RecrelTr0}) with $j=i+2$ ensures 
that $t_{i,i+2}(p)=0$ too. At this point it is crucial that the coefficient 
$B_{(0,2)(i,i+2)}(p)=0$. That $t_{i,j}(p)=0$ also for $j>i+2$ now simply follows 
from the structure of the recursion relation (\ref{RecrelTr0}). 

To conclude, we have shown that 
our statement\\
(a) is correct for $i'=0$\\
(b) the statement for $i'=i$ follows from those with $i'<i$\\
thus completing the proof.

\section{Discussion}

In \cite{Poghosyan:2025rhj}, the recursion relation~\eqref{RecrelTr0} was derived. 
By iterating this relation starting from the initial condition~\eqref{ini_cond_00}, 
it was observed that \( t_{i,j} = 0 \) whenever \( j > i \). 
This empirical observation led to the conjecture that the 5-point conformal block 
with a level-two degenerate field insertion has the form given in 
\eqref{Ganz1}, \eqref{H1}, and \eqref{Pfin}. 
The crucial point concerns the structure of \( P(x,z) \) and \( \hat{P}(x,z) \), 
as defined in~\eqref{Pfin}, where the summation over \( j \) runs from \( 0 \) to \( i \), 
rather than from \( 0 \) to \( \infty \). 

Although the recursion relation can be implemented efficiently---for instance in \textit{Mathematica}---and provides strong evidence 
that the property \( t_{i,j} = 0 \) for \( j > i \) persists even for very large values 
of \( i \) and \( j \), such numerical verification cannot exclude the 
possibility of a breakdown at some arbitrarily high order. 
The purpose of the present paper is to remove this uncertainty by providing a 
rigorous proof. 

The vanishing of \( t_{i,j} \) for \( j > i \) is in fact essential: at any fixed 
order in \( x \), it implies that both \( P \) and \( \hat{P} \) are \emph{finite} 
polynomials in \( z^{-1} \). 
This, in turn, ensures that our representation of the conformal block is exact in \( z \) at each fixed order in \( x \). 
It is precisely this property that allows one to analytically continue to the region \( z \gg 1 \) 
and get the four-point block via the OPE with the field at infinity; see \cite{Poghosyan:2025rhj}.

\section*{Acknowledgements}
The research of R.P. was supported by the Armenian SCS grants 21AG-1C060 and 24WS1C031. Similarly, H.P.’s work received support from Armenian SCS grants 21AG-1C062 and 24WS-1C031.




\bibliographystyle{elsarticle-num} 



\end{document}